# SimBins: An information-theoretic approach to link prediction in real multiplex networks

Seyed Hossein Jafari[1*], Amir Mahdi Abdolhosseini-Qomi[1], Masoud Asadpour[1], Maseud Rahgozar[1], Naser Yazdani[1]

[1] School of Electrical and Computer Engineering, College of Engineering, University of Tehran, Tehran, Iran

* Corresponding author
E-mail: jafari.h@ut.ac.ir (SHJ)

# Abstract

The entities of real-world networks are connected via different types of connections (i.e. layers). The task of link prediction in multiplex networks is about finding missing connections based on both intra-layer and inter-layer correlations. Our observations confirm that that in a wide range of real-world multiplex networks, from social to biological and technological, a positive correlation exists between connection probability in one layer and similarity in other layers. Accordingly, a similarity-based automatic general-purpose multiplex link prediction method –SimBins– is devised that quantifies the amount of connection uncertainty based on observed inter-layer correlations in a multiplex network. Moreover, SimBins enhances the prediction quality in the target layer by incorporating the effect of link overlap across layers.

Applied to various datasets from different domains, SimBins proves to be robust and superior than compared methods in majority of experimented cases in terms of accuracy of link prediction. Furthermore, it is discussed that SimBins imposes minor computational overhead to the base similarity measures making it a potentially fast method, suitable for large-scale multiplex networks.

# Introduction

Link prediction has been an area of interest in the research of complex networks for over two decades [1], studying the relationships between entities (nodes) in data represented as graphs. The main goal is to reveal the underlying truth behind emerging or missing connections between node pairs of a network. Link prediction methods have a wide range of applications, from discovery of latent and spurious interactions in biological networks (which is basically quite costly if performed in traditional methods) [2, 3] to recommender systems [4, 5] and better routing in wireless mobile networks [6]. Numerous perspectives have been adopted to attack the problem of link prediction.

Similarity-based methods tend to measure how similar nodes are as an indication of likelihood of linkage between them. This approach is a result of assuming two nodes are similar if they share many common features [7]. A whole lot of nodes' features stay hidden (or are kept hidden intentionally) in real networks. Additionally, it is an interesting question that despite of hiding a considerable amount of network information, what fraction of the truth behind a process (e.g. link formation) can still be extracted by solely including *structural features*? That is one of the main drives to utilize structural similarity indices for link prediction. Several different classifications of similarity measures have been proposed, among all, classifying based on locality of indices is of great importance. To name a few, Common Neighbors (CN) [1], Preferential Attachment (PA) [8], Adamic-Adar (AA) [9] and Resource Allocation (RA) [10] are popular indices focusing mostly on nodes' structural features, each with unique characteristics. Despite their simplicity, these indices are popular due to their low computational cost and reasonable prediction performance. On the other hand, global indices take features of the whole network structure into account, tolerating higher cost of computation, usually in favor of more accurate information. Take length of paths between pairs of nodes for instance, which the well-known Katz [11] index operates on. Average Commute Time (ACT) [1] and PageRank [12] are some other notable global indices. Somewhere in between lies the quasi local methods which inherit properties from both local and global indices meaning that although they utilize some global network information, computational complexity is kept comparable to local methods, such as the Local Path (LP) [13] index and Local Random Walk (LRW) [14]. For more detailed information on these similarity indices (also described as *unsupervised* methods in the literature [15]), readers are advised to refer to [16].

Some researchers have tackled the link prediction problem using the ideas of information theory; as in [17] mutual information (MI) of common neighbors is incorporated to estimate the connection likelihood of a node pair. Moreover, Path Entropy (PE) [18] similarity index has been conducted which not only takes quantity and length of paths between a pair of nodes into account, but also considers the entropy of those paths affecting connection likelihood of the pair.

From a coarse-grained point of view, *supervised* models of link prediction reside in a different class than aforementioned unsupervised ones. They learn a group of parameters by processing input graph and use certain models, such as feature-based prediction (HPLP [19]) and latent feature extraction (Matrix Factorization [15]). Representation learning has helped automating the whole process of link prediction especially feature selection, one such example method is node2vec [20]. Learning-based methods usually lead to better results compared to similarity-based counterparts, but this does not mean that unsupervised models should be considered obsolete. On the one hand, unsupervised models provide a clearer insight on underlying characteristics of networks, take common neighbors (CN) for example which indicates the high clustering property of networks [18] or Adamic-Adar index which is based on the size of common nodes' neighborhoods [9]. On the other hand, unsupervised methods can take much less computation effort, which makes them suitable for online prediction without any high cost training phase or feature selection process [21].

## Related Works

As said so far, complex networks research was focused on single-layer networks (simplex or monoplex) for many years. The study of multi-layer (multiplex or heterogeneous) networks has gained the attention of researchers in the past few years. Refs. [22, 23] provide noteworthy reviews on history of multi-layer networks. Attempts for multi-layer link prediction are not abundant in which some of them are introduced here.

Hidden geometric correlation in real multiplex networks [24] is an interesting work which depicts how multiplex networks are not just random combinations of single-layer networks. They employ these geometric correlations for trans-layer link prediction i.e. incorporating observations of other layers for predicting connections in a specific layer. This work is followed by a study that argues the requirement of a link persistence factor to explain high edge overlap in real multiplex systems [25]. In heterogeneous networks (i.e. networks with different types of nodes and relations), several similarity-search approaches have been proposed. PathSim [26] is a meta path-based similarity measure that can find similar peers in heterogeneous networks (e.g. authors in similar fields in a bibliographic network). The intuition behind PathSim is that two peer objects are similar if they are not only strongly connected, but also share comparable visibility (number of path instances from a node to itself). HeteSim [27] is another method of the same kind which can measure similarity of objects of different type, inspired by the intuition that two objects are related if they are referenced by related objects. Their drawback, however, is their dependence on connectivity degrees of node-pairs (neglecting further information provided by meta paths themselves) and their necessity of using one and usually symmetric meta-path. In [28], a mutual information model has been employed to tackle these problems. Most meta path-based models suffer from lack of automated meta-path selection mechanism, in other words, pre-defined meta paths (mostly specific to the dataset under study) are utilized to help with prediction tasks. Another major issue of previously discussed methods is that by including longer meta paths, much more computation is needed to analyze these paths and their role in prediction.

Some researchers have approached the problem of link prediction in multiplex networks using feature engineering and application of machine learning. A study of a multiplex online social network, demonstrates the importance of multiplex links (link overlap) in significantly higher interaction of users based on available side information [29]. The authors consider Jaccard similarity of extended neighborhood of nodes in the multiplex network as a feature for training a classifier for link prediction task. They have shown that using this multiplex feature enhances the link prediction performance. A similar work on the same dataset benefits from node-based and meta-path-based features [30]. A specialized type of these meta-paths is tailored to be originated from and ending at communities. The effectiveness of the features has been examined by a binary classification for link predication task. Recently, other interlayer similarity features, based on degree, betweenness, clustering coefficient and similarity of neighbors has been used [31].

Furthermore, the issue of link prediction has been investigated in a scientific collaboration multiplex network [32]. The authors have proposed a supervised rank aggregation paradigm to benefit from the node pairs ranking information which is available in other layers of the network. Another study uses rank aggregation method on a time-varying multiplex network [33]. The effect of other layers on the target layer of link prediction has been measured using global link overlap. A recent work combines feature engineering and rank aggregation [34]. Two features based on hyperbolic distance are being used and link overlap is considered for relevance of layers.

The issue of layer relevance and its effect on link prediction is studied in [35]. The authors use global link overlap and Pearson correlation coefficient of node features as measures of layer relevance and later they use it to combine the basic similarity measures of each layer. The results support that the more layers are relevant, the better performance of link prediction is attained.

A systematic approach is extending the basic similarity measures to multiplex networks. However, when it comes to multiplex networks, it's hard to extend the notion of similarity [36]. In this paper, an information-theoretic model is devised that employs other layers' structural information for better link prediction in some arbitrary (target) layer of the network. By incorporating several similarity indices (RA, CN and ACT) as base proximity measures, we will describe that the proposed method -SimBins- can be used to extend all similarity indices for multiplex link prediction without significantly degrading time complexity. Finally, it is shown that SimBins improves prediction performance on several different real-world social, biological and technological multiplex networks.

# Materials and Methods

## Link Prediction in Multiplex Networks

Consider a multiplex network $G\left(V, E^{[1]}, ..., E^{[M]}; E^{[\alpha]} \subseteq V \times V \quad \forall \alpha \in \{1, 2, ..., M\}\right)$ where $M$, $V$ and $E^{\alpha}$ are the number of layers, the set of all nodes and existing edges in layer $\alpha$ of the multiplex network, respectively. Let $U = V \times V$ be the set of all possible node pairs. Current research aims to study undirected multiplex networks; therefore, it is assumed that $G(V, E^{\alpha})$ for any arbitrary layer $\alpha$ is an undirected simple graph. The link prediction in multiplex networks is concerned with the issue of predicting missing links in an arbitrary target layer $T \in \{1, 2, ..., M\}$ with the help of other auxiliary layers. To be able to evaluate the proposed method, $E^T$ i.e. the edges in target layer is divided into a training set $E^T_{train}$ ( 90% of $E^T$ ) and a test set $E^T_{test}$ ( 10% of $E^T$ ) so that $E^T_{train} \bigcup E^T_{test} = E^T$ and $E^T_{train} \bigcap E^T_{test} = \varnothing$. Only the information provided by the training set is used in the prediction task and eventually, $E^T_{test}$ is compared to the output of the proposed algorithm (link-existence likelihood scores for a subset of $U - E^T_{train}$, including $E^T_{test}$), determining the performance of the method. To be more specific, link likelihood scores are calculated for node pairs of $E^T_{test}$ and a random subset $Z^T_{test}$ of $U - E^T$ where $|Z^T_{test}| = 2|E^T_{test}|$ for which all of them are disconnected in $E^T_{train}$. To put it in a few words; only a subset of non-observed links in training set are scored for the sake of complexity which will be discussed in detail later. Notice the coefficient $2$, which is a ratio incorporated to satisfy the link imbalance assumption in real-world networks (that are mostly sparse by nature [37]).

In the present study, the issue under scrutiny is how employing one layer of the multiplex network such as $A$, facilitates the task of link prediction in another layer $T$ where $T, A \in \{1,...,M\}; \ T \neq A$ i.e. a *duplex* subset of the multiplex network. In 'Discussion' section, it is argued that how one can extend the proposed method to utilize the structural information of multiple layers for link prediction.

## Evaluation Method

In their ideal form, link prediction algorithms tend to rank non-observed links in a network so that all latent links are situated on top of the ranking and all other non-existent links underneath. This ranking is based on a link-likelihood score that is dedicated to node pairs corresponding to non-observed links in the network. Acquisition of Area Under Receiver Operating Characteristic Curve (AUC or AUROC) [38] is prominent in the literature for evaluating link prediction methods [16]. AUC indicates the probability that a randomly chosen missing link is scored higher than a randomly chosen non-existent link, denoted as:

$$\text{AUC} = \frac{n' + 0.5 n''}{n} \quad (1)$$

where by performing $n$ times of independent comparisons ($n = 10000$ in our experiments), a randomly chosen latent link has a higher score compared to a randomly chosen non-existent link in $n'$ times and are equally scored in $n''$ times. AUC will be $1$ if the node pairs are flawlessly ranked and $0.5$ if the scores follow an identical and independent distribution i.e. the higher the AUC, the better the scoring scheme is.

## Data

Various real-world multiplex network datasets from different domains are selected for investigation; from social (Physicians, NTN and CS-Aarhus) to technological (Air/Train and London Transport) and biological systems (C. Elegans, Drosophila and Human Brain). They also have diverse characteristics that are briefly introduced in Table 1.

- *Air/Train (AT).* This dataset consists of Indian airports network and train stations network and their geographical distances [39]. To relate the **train** stations to the geographically nearby **air**ports, in [24] they have aggregated all train stations within 50km from an airport into a super-node. Then, the super-nodes are considered as connected if they share a common train station, or if one train station of one super-node is directly connected to a station of the other super-node. Air is the network of airports and Train is the network of aggregated train station super-nodes.
- *C. Elegans.* The network of neurons of the nematode Caenorhabditis Elegans that are connected through miscellaneous synaptic connection types: **Electric**, **Chemical Monadic** and **Chemical Polyadic** [40].
- *Drosophila Melanogaster (DM).* Layers of this network represent different types of protein-protein interactions belonged to the fly Drosophila Melanogaster, namely **suppressive** genetic interaction and **additive** genetic interaction. More details can be found in [41, 42].
- **Human Brain (HB).** The human brain multiplex network is taken from [24, 43]. It consists of a **structural** or anatomical layer and a **functional** layer that connect 90 different regions of the human brain (nodes) to each other. The structural network is gathered by dMRI and the functional network by BOLD fMRI [43]. In this multiplex network, the structural connections are obtained by setting a threshold on connection probability of brain regions (which is proportional to density of axonal fibers in between) [24]. The functional interactions are derived in a similar manner, by putting a threshold on the connection probability of regions which is proportional to a correlation coefficient measured for activity of brain region pairs [24].
- *Physicians.* Taken from [44], the Physicians multiplex dataset contains 3 layers which relate physicians in four US towns by different types of relationships; to be specific, **advice**, **discuss** and **friendship** connections.
- *Noordin Top Terrorist Network (NTN).* Taken from [45], this multiplex dataset is made of information among 78 individuals i.e. Indonesian terrorists that depicts their relationships with respect to exchanged **communications**, **financial** businesses, common **operations** and mutual **trust**.
- *London Transport.* For the purpose of studying navigability performance under network failures, De Domenico et al. [46] gathered a dataset for public transport of London consisting of 3 different layers; the **tube**, the **overground**, and the docklands light railway (**DLR**). Nodes are stations which are linked to each other if a real connection exists between them in the corresponding layer.
- *CS-Aarhus.* This dataset is collected from [47] which is conducted at the Department of Computer Science at Aarhus University in Denmark among the employees. The network consists of 5 different interactions corresponding to current **work** relationships, repeated **leisure** activities, regularly eating **lunch** together, **co-author**ship of publications and friendship on **Facebook**.

Node multiplexity in Table 1 shows the fraction of nodes in a multiplex network that are active (have at least one link attached) in more than one layer.

Table 1 Basic Characteristics of Multiplex Networks Used in Experiments.

| MULTIPLEX NAME | # LAYERS | # NODES | NODE MULTIPLEXITY | LAYER NAME | # ACTIVE NODES | # LINKS |
|---|---|---|---|---|---|---|
| Air/Train | 2 | 69 | 1 | Air | 69 | 180 |
| | | | | Train | 69 | 322 |
| C. Elegans | 3 | 280 | 0.98 | Electric | 253 | 515 |
| | | | | Chem-mono | 260 | 888 |
| | | | | Chem-poly | 278 | 1703 |
| Drosophila | 2 | 839 | 0.89 | Suppress | 838 | 1858 |
| | | | | Additive | 755 | 1424 |
| Brain | 2 | 90 | 0.85 | Structure | 85 | 230 |
| | | | | Function | 80 | 219 |
| Physicians | 3 | 246 | 0.93 | Advice | 215 | 449 |
| | | | | Discuss | 231 | 498 |
| | | | | Friend | 228 | 423 |
| NTN | 4 | 78 | 0.94 | Communication | 74 | 200 |
| | | | | Financial | 13 | 15 |
| | | | | Operational | 68 | 437 |
| | | | | Trust | 70 | 259 |
| London | 3 | 368 | 0.13 | Tube | 271 | 312 |
| | | | | Overground | 83 | 83 |
| | | | | DLR | 45 | 46 |
| CS-Aarhus | 5 | 61 | 0.96 | Lunch | 60 | 193 |
| | | | | Facebook | 32 | 124 |
| | | | | Co-author | 25 | 21 |
| | | | | Leisure | 47 | 88 |
| | | | | Work | 60 | 194 |

## Information Theory Background

This sub-section is concerned with the issue of introducing necessary concepts of information theory, as it lays out the main mathematical background of the proposed method. What follows is the definition of self-information and mutual information.

Given a random variable $X$, the *self-information* or surprisal of occurrence of event $x \in X$ with probability $p(x)$ is defined as [48]:

$$I(X = x) = -\log p(x) \qquad (2)$$

The self-information implies how much uncertainty or surprise there is in the occurrence of an event; the less probable the outcome is, the more the surprise it conveys. The base of the logarithmic functions is assumed to be $2$ throughout the paper, as they measure uncertainty in *bits* of information.

Let's proceed with the definition of mutual information between two random variables $X$ and $Y$ with joint probability mass function $p(x,y)$ and marginal probability mass functions $p(x)$ and $p(y)$, respectively. The *mutual information* $I(X;Y)$ is [49]:

$$\begin{aligned} I(X;Y) &= \sum_{x \in X} \sum_{y \in Y} p(x,y) \log \frac{p(x,y)}{p(x)p(y)} \\ &= \sum_{x,y} p(x,y) \log \frac{p(x,y)}{p(x)p(y)} \\ &= \sum_{x,y} p(x,y) \log \frac{p(x|y)}{p(x)} \end{aligned} \qquad (3)$$

Consequently, the mutual information of two events $x \in X$ and $y \in Y$ can be denoted as [17]:

$$I(X = x; Y = y) = \log \frac{p(x \mid y)}{p(x)} = -\log p(x \mid y) - (-\log p(x)) \quad (4)$$
$$= I(x) - I(x \mid y)$$

In fact, the mutual information indicates how much two variables are dependent to each other i.e. for a variable $X$, how much uncertainty is reduced due to observation of another variable $Y$. The mutual information would be zero if and only if two variables are independent. In the following section, we will describe how these two measures play their roles in designation of our method.

## Base Similarity Measures

There is extensive literature on similarity measures that determine how similar two nodes are in a single-layer network; as it was partially presented on introduction of this paper. In our proposed method, a subset of these similarity indices (both local and global) is used as base measures that the multiplex link prediction model is built on top of them.

- **CN** [1]: Maybe, the most well-known and typical way to measure similarity of two nodes $x$ and $y$ is to count the number of their common neighbors:
$$S_{xy}^{CN} = |\Gamma(x) \cap \Gamma(y)| \quad (5)$$
where $\Gamma(x)$ and $\Gamma(y)$ are the set of neighbors of $x$ and $y$, respectively.

- **RA** [10]: In Resource Allocation, degree of a node is considered as a resource that is allocated to the neighbors of that node negatively proportional to its degree:
$$S_{xy}^{RA} = \sum_{z \in \Gamma(x) \cap \Gamma(y)} |\Gamma(z)|^{-1} \quad (6)$$

- **ACT** [1]: Random-walk based methods account for the steps required for reaching one node starting from some arbitrary node. Average Commute Time measures the average number of steps required for a random walker to reach node $y$ starting from node $x$. For the sake of computational complexity, pseudo-inverse of Laplacian matrix is utilized to calculate the commute time:
$$S_{xy}^{ACT} = \frac{1}{l_{xx}^+ + l_{yy}^+ - 2l_{xy}^+} \quad (7)$$
where $l_{xy}^+$ is the $[x, y]$ entry in pseudo-inverse Laplacian matrix i.e. $l_{xy}^+ = [L^+]_{xy}$. The pseudo-inverse of Laplacian is calculated as [50]:
$$L^+ = \left(L - \frac{ee'}{n}\right)^{-1} + \frac{ee'}{n} \quad (8)$$
where $e$ is a column vector of 1's ($e'$ is its transpose) and $n$ is the total number of nodes.

## Results

Does the structure of one layer of a multiplex, provide any information on the formation of links in some other layer of the same network? Take a social multiplex network, for example, in which one layer states people's work relationships and the other layer represents their friendship. Intuitively it can be conjectured that in a real multiplex like our sample social network, structural changes in one layer can affect the others; if two people become colleagues, the conditions of them being friends will probably not be the same as it was before. More specifically, is there any correlation among the structure of layers of a multiplex network? This question has been positively answered in previous studies with different approaches. In [24] a null model is created for a multiplex network, by randomly reshuffling inter-layer node-to-node mappings. Subsequently, it is shown that geometric inter-layer correlations are destroyed in the null model compared to the original network.

Various structural features can be analyzed to uncover correlations between layers. Direct links, common neighbors, paths [1] and eigenvectors [51] are such examples. In the following sections we will develop a set of tools that assist in collection of evidences about inter-layer correlations in multiplex networks, as basic intuitions supporting the proposed link prediction framework.

## Partitioning Node Pairs (Binning)

Consider two layers $T, A \in \{1, 2, ..., M\}; T \neq A$ of a multiplex network with $M$ layers and $V$ nodes. $T$ is the target layer, so it is intended to predict likelihood of presence of links in that layer, and $A$ is the auxiliary layer assisting the prediction task. A subset $U'$ of $U = V \times V$ is constituted so that $U' = E^T_{train} \cup Z^T_{train}$ where $Z^T_{train}$ is a random sample of non-observed links from $U - E^T$ and $|Z^T_{train}| = 2|E^T_{train}|$. The size of $Z^T_{train}$ is twice as large as $E^T_{train}$, so that $U'$ would be a suitable representative of the target layer due to the link imbalance phenomenon in real complex systems. Two different partitions of $U'$ is formed (using equal-depth binning, described in the following paragraph):

(i) w.r.t the target layer $T$:

$$\{S^T_1, S^T_2, ..., S^T_{b_T}\} \text{ where } \bigcup_{i=1}^{b_T} S^T_i = U' \text{ and } \forall i, j \in \{1, 2, ..., b_T\}, i \neq j \Rightarrow S^T_i \cap S^T_j = \emptyset$$

(ii) With respect to the auxiliary layer $A$:

$$\{S^A_1, S^A_2, ..., S^A_{b_A}\} \text{ where } \bigcup_{j=1}^{b_A} S^A_j = U' \text{ and } \forall i, j \in \{1, 2, ..., b_A\}, i \neq j \Rightarrow S^A_i \cap S^A_j = \emptyset$$

These partitions are introduced as **bins** of node pairs in current study. The number of bins w.r.t target and auxiliary layer are $b_T$ and $b_A$, respectively. An equal-depth (frequency) binning strategy is applied to the target layer similarity scores of the node pairs in $U'$, in order that each partition $S^T_i; i \in \{1, 2, ..., b_T\}$ contains approximately the same number of members (node pairs). The same strategy goes for similarity scores in auxiliary layer $A$, establishing $S^A_j; j \in \{1, 2, ..., b_A\}$ partitions. Aforementioned partitions (bins) form the building blocks of how the multiplex networks are scrutinized in this paper, as they put forward a coarse-grained view of the data; tolerating the statistically insignificant phenomena observed in particular regions of the networks. The setting denoted above will be used from now onwards, to avoid any further repetitions.

## Intra-layer and Trans-layer Connection Probabilities

The foregoing discussion introduces two key measures for target and auxiliary layer bins, namely $S^T_i$ and $S^A_j$: (1) intra-layer connection probability $p_{intra}(S^T_i)$, and (2) trans-layer connection probability $p^T_{trans}(S^A_j)$. Intra-layer connection probability in $S^T_i$ is the connection likelihood of pairs existing in that bin. This measure can also be expressed as conditional probability of connection of an arbitrary node pair $x, y$ in layer $T$, given their similarity (bin) in the same layer:

$$p_{intra}(S^T_i) = p(L^T = 1 | S^T_i); \quad i \in \{1, 2, ..., b_T\} \tag{9}$$

Notice $L^T = 1$, which is the event that any randomly selected pair $(x,y)$ are linked in layer $T$. Empirically, $p_{\text{intra}}(S_i^T)$ is computed as proportion of linked node pairs in $S_i^T$ to all of node pairs in the set:

$$\tilde{p}_{\text{intra}}(S_i^T) = \frac{|S_i^T \cap E_{\text{train}}^T|}{|S_i^T|}; \quad i \in \{1, 2, ..., b_T\} \tag{10}$$

Intra-layer connection probability for four different multiplex (duplex) networks is provided for each bin in (Fig 1). In data-driven observations of this paper, wherever a similarity measure is involved, Resource Allocation (RA) index is used; otherwise specified. Additionally, it is assumed that the number of bins in both the target and auxiliary layers i.e. $b_T$ and $b_A$ are set to $10$. Our experiments show that too small number of bins leads to significant decrement in prediction results.

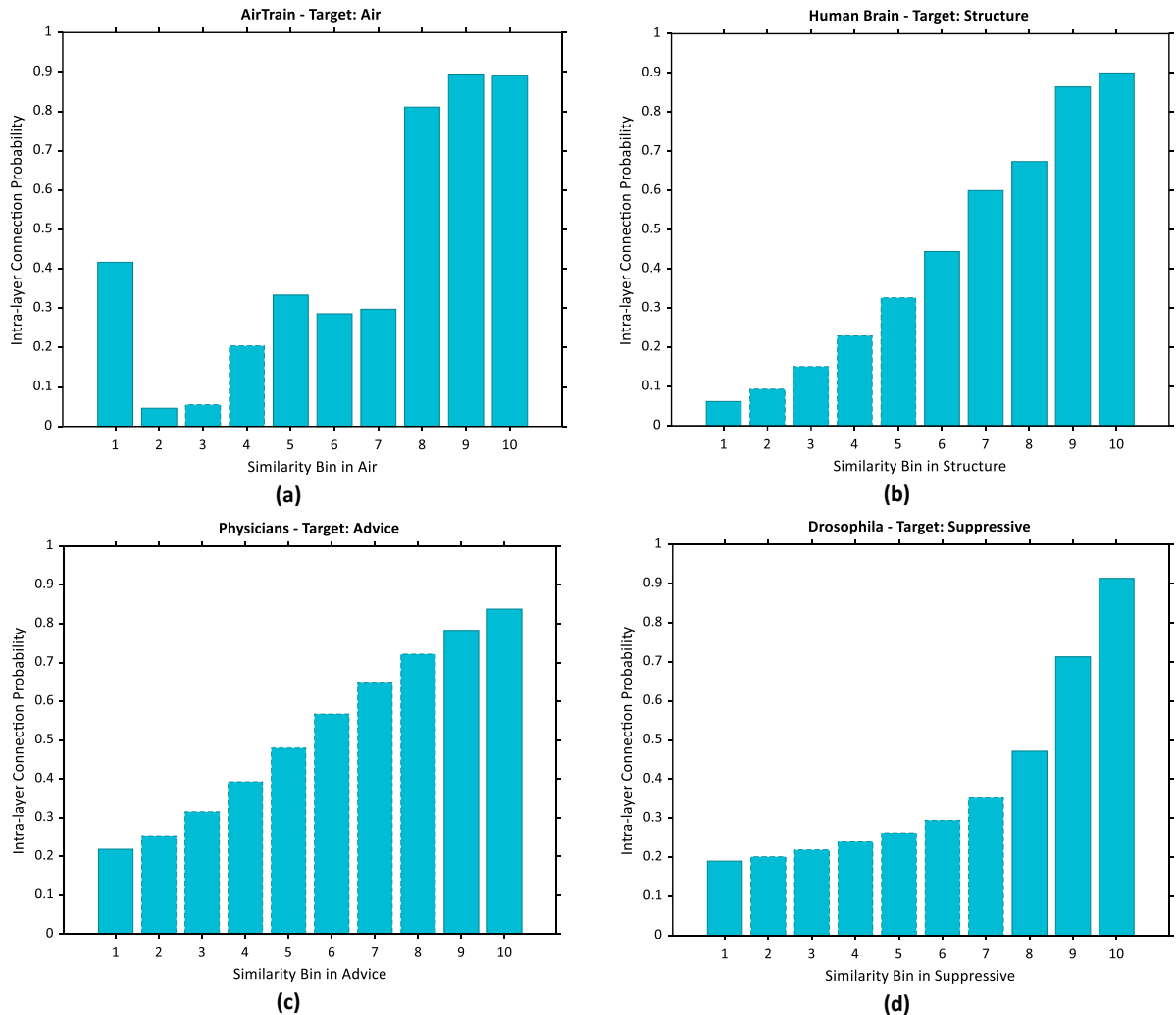

**Fig 1 Intra-layer connection probability in target layer bins.** Intra-layer connection probability or fraction of node pairs in a bin that are linked in layer (a) 'Air' of the network Air/Train, (b) 'Structure' of Human Brain, (c) 'Advice' of Physicians, (d) 'Suppressive' of Drosophila. Bars with dashed lines represent imputed probabilities.

In most of the cases, increasing the number of bins either has no effect on prediction results or degrades them (although not quite significantly). Additionally, large number of bins brings unnecessary computational complexity to our algorithm. We have also tried a more adaptive approach for choosing the number of bins by maximizing the entropy of node-pairs distribution in bins which lead to no substantial improvement in prediction. A value between 10 and 50 is recommended as SimBins shows no significant sensitivity in terms of accuracy within the mentioned range and the computational overhead is miniscule.

The bars with dashed lines in (Fig 1) represent imputed values. Because of high frequency of some certain similarity values (such as 0 scores in RA for node pairs with no common neighbors), a perfect equal-depth binning may not be feasible; as a result, a number of bins will contain no sample node pairs. The value of intra-layer connection probability for these bins has been imputed using a penalized least squares method which allows fast smoothing of gridded (missing) data [52]. In addition to more clear observations, this imputation will let us fix the number of bins and handle missing data in a systematic way. The results indicate that by the increment of similarity (higher bin numbers) intra-layer connection probability increases respectively, depicting a positive correlation between similarity (bin number) and intra-layer connection probability; as stated in seminal work of Liben-nowell and Kleinberg [1].

Trans-layer connection probability is defined analogously except that although connection in target layer $T$ is concerned, the similarity scores of node pairs are given in auxiliary layer $A$. Similar to formula (9), $p_{\text{trans}}^T(S_j^A)$ can be defined as follows:

$$p_{\text{trans}}^T(S_j^A) = p(L^T = 1 | S_j^A); \quad j \in \{1, 2, ..., b_A\} \tag{11}$$

Empirical value of trans-layer connection probability is calculated likewise:

$$\tilde{p}_{\text{trans}}^T(S_j^A) = \frac{|S_j^A \cap E_{\text{train}}^T|}{|S_j^A|}; \quad j \in \{1, 2, ..., b_A\} \tag{12}$$

In other words, $p_{\text{trans}}^T$ w.r.t $A$ relates the similarity of node pairs in layer $A$ to their probability of connection in layer $T$. Trans-layer connection probability of four duplexes is depicted in the left column of (Fig 2). Moreover, the node pairs in $S_j^A$ can be divided into two disjoint sets based on their connectivity in the auxiliary layer. Then the trans-layer connection probability for connected node pairs in auxiliary layer $S_j^A \cap E^A$ and unconnected ones $S_j^A \cap (U - E^A)$ will be

$$\tilde{p}_{\text{trans}}^T(S_j^A \cap E^A) \tag{13}$$

and

$$\tilde{p}_{\text{trans}}^T\left(S_j^A \cap (U - E^A)\right) \tag{14}$$

as shown in the middle and right columns of (Fig 2), respectively.

The bars with dotted lines represent imputed trans-layer connection probabilities, similar to intra-layer connection probabilities in (Fig 1). By inspecting the values of trans-layer connection probabilities for the datasets under study, a rising pattern is prominent by moving to bins corresponding to higher similarity ranges. Drosophila in (Fig 2. d1-3) brings up an exceptional case, where similarity in the auxiliary (Additive) layer shows no correlation with connection in the target (Suppressive) layer. Except these kind of irregularities in data, the available evidence appears to suggest that in most of the real multiplex networks, probability of connection in one (target) layer of the network does have positive correlation with similarity in some other (auxiliary) layer i.e. as similarity grows higher in the auxiliary layer, it can be a signal of higher connection probability in target layer. This observation develops the claim that for link prediction in target layer, not only the similarity of nodes in that same layer, but also their similarity in some other auxiliary layer can be utilized. Notice that this rising pattern in $p_{\text{trans}}$ is observed in almost all datasets under scrutiny, independent from the choice of similarity measure.

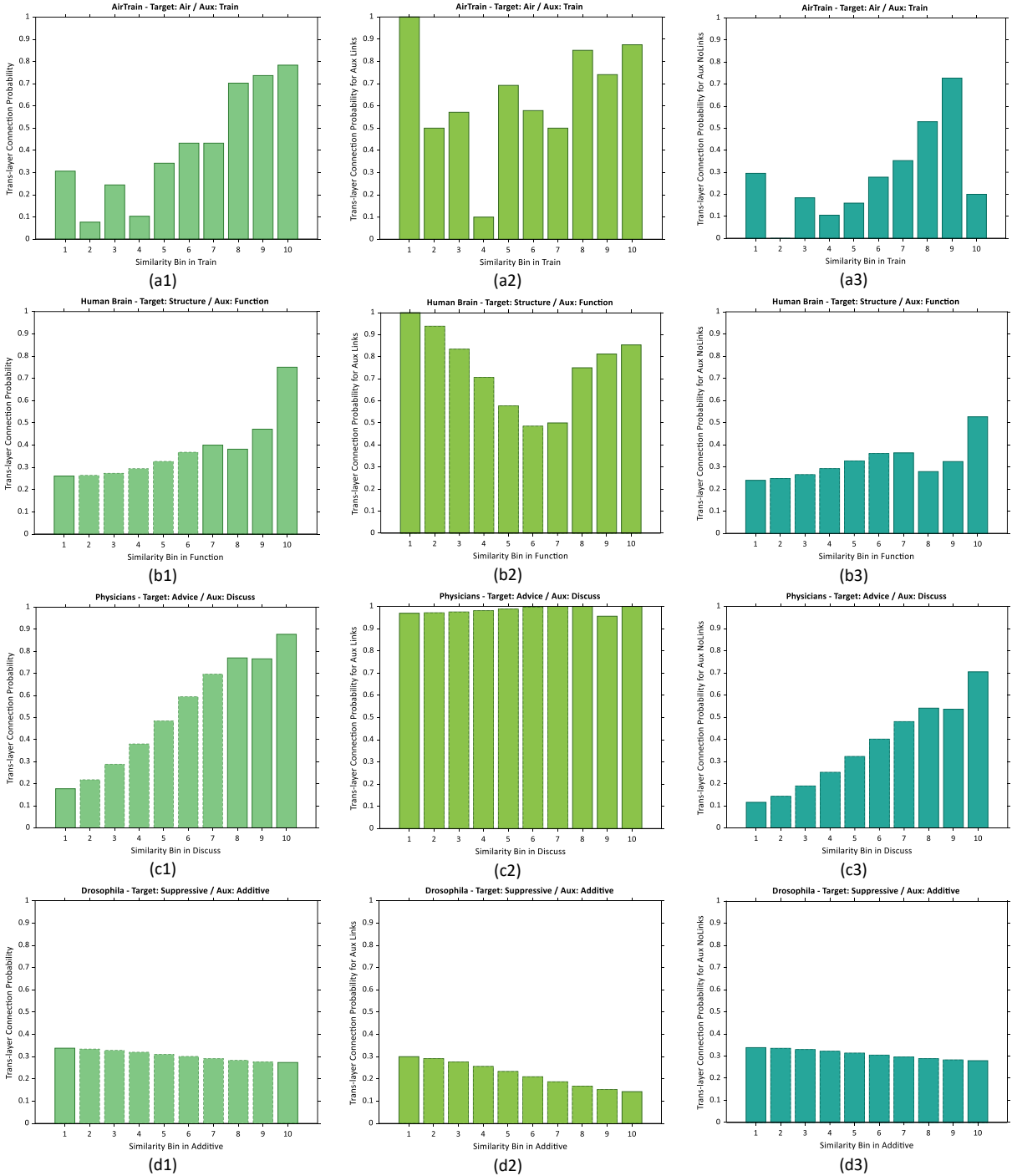

**Fig 2 Empirical trans-layer connection probability in auxiliary layer bins.** (a1-d1) Trans-layer connection probability of all node pairs, (a2-d2) Trans-layer connection probability of node-pairs connected in auxiliary layer, (a3-d3) Trans-layer connection probability of node-pairs unconnected in auxiliary layer, for sample duplexes of 4 datasets

The previously described property of trans-layer connection probability lies at the heart of the current study, shaping the main idea of the proposed multiplex link prediction method. In addition, the connectedness of the node pairs in the auxiliary layer leads to significant increase in the trans-layer connection probabilities. In Human Brain and Physicians networks the presence of link in the auxiliary is a strong evidence of connectivity in the target layer. The case is similar for AirTrain network but with lower certainty. The Drosophila network is an exception as before. These findings are in consistence with the link persistence phenomenon as reported in [25]. Here, we propose a consolidated method which considers the similarity of node pairs in the target and auxiliary layers, and also their connectedness in the auxiliary layer as the underlying evidences for calculating the uncertainty of linkage in the target layer.

Furthermore, by simultaneously partitioning $U'$ based on their similarity in both target and auxiliary layers, we obtain $b_T \times b_A$ partitions or *2d-bins*. Within each 2d-bin, the fraction of target layer links to total node pairs is included i.e. the empirical connection probability in target layer is computed. In (Fig 3), empirical probability of connection in 2d-bins is presented for the same duplexes as in (Fig 2).

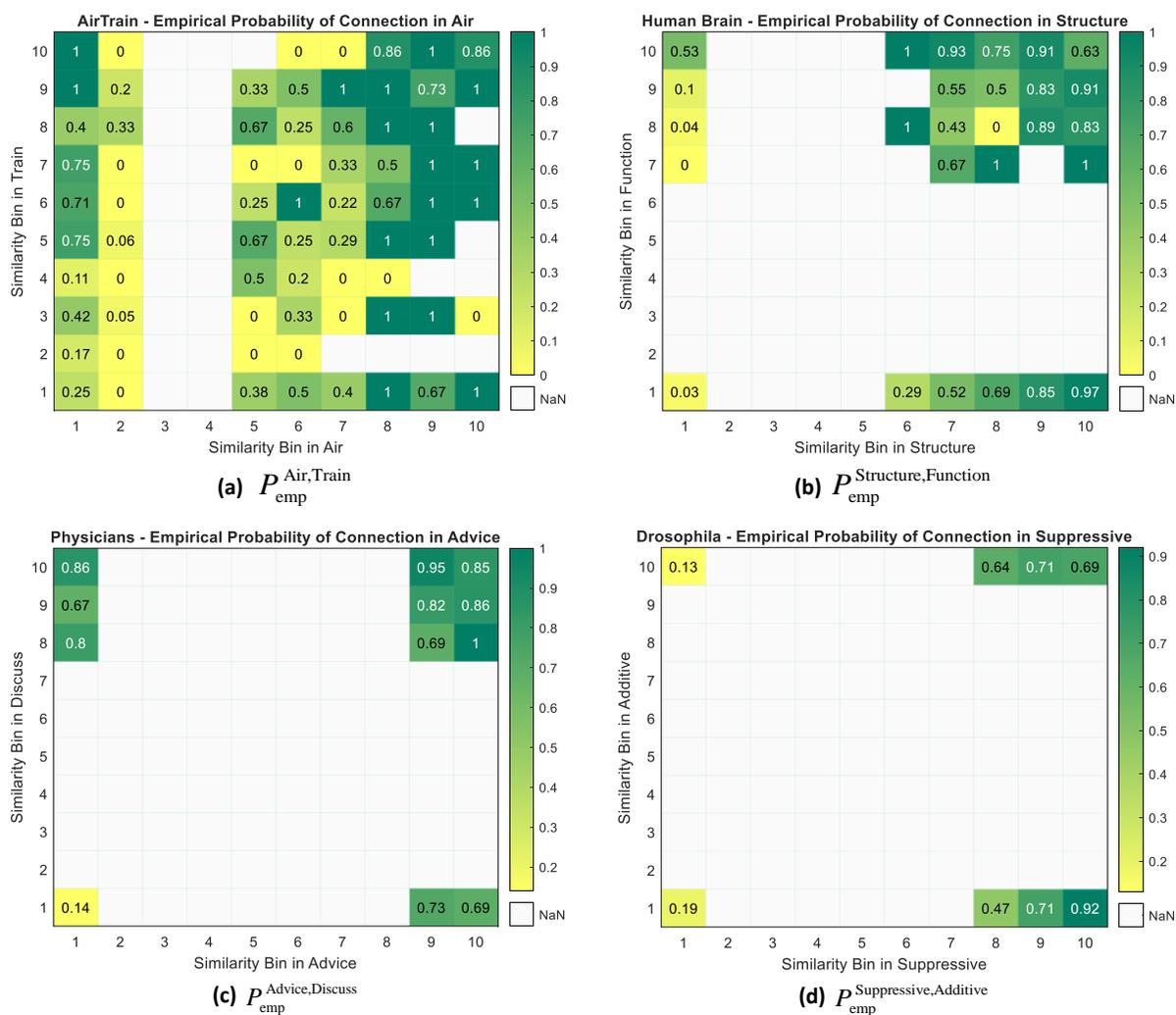

**Fig 3 Empirical probability of connection in 2d-bins.** The fraction of node pairs in the 2d-bins that are connected in the target layer(a) 'Train' of the network Air/Train w.r.t 'Air', (b) 'Function' of Human Brain w.r.t 'Structure', (c) 'Discuss' of Physicians w.r.t 'Advice', (d) 'Additive' of Drosophila w.r.t 'Suppressive' layer. NaN (Not a Number) values represent 2d-bins that contain no sample pairs.

Several results can be inferred by scrutinizing (Fig 3). Increment of the empirical probability of connection in the horizontal axis expresses the effectiveness of the similarity measure in target layer; the higher the bin number, the larger the fraction of node pairs that have formed links. Another aspect of the above figure is the ascension of the empirical probability of connection by moving to higher bin number in the auxiliary layer i.e. the vertical axis (except Drosophila in Fig 3. d1-3), which is a sign of positive correlation between the probability of connection in target layer and similarity in the auxiliary layer; so far totally consistent with Fig 1 and Fig 2. This cross-layer connection and similarity correlation are observed in the majority of datasets under study, in which a subset of them are presented above. It is interesting that when similarity of a node-pair is very low in the target layer, high similarity in the auxiliary layer leads to stronger connection probability between them.

A subtle observation on the data comprises a difference in the ascension pace of empirical connection probability in target similarity (horizontal axis) versus auxiliary similarity (vertical axis). Based on the evidence currently available, it seems fair to suggest that, although the growth of trans-layer connection probability increases the empirical probability of connection in the target layer, intra-layer similarity brings it up faster. It can be deduced that intra-layer similarities play more important roles compared to trans-layer similarities. Therefore, later in the proposed model, the intra-layer connection probability will be considered a stronger signal than the trans-layer counterpart.

The following sub-sections are concerned with the issue of how to estimate probability of connection in the target layer of a multiplex network by incorporating other layers' structural information with a systematic approach that generalizes beyond specific data.

## Fusion of Decisions

Consider two independent decision makers that determine the probability of occurrence of a certain event corresponding to a binary random variable. Each of them declares a probability $p$ and $q$ (where $0 \leq p, q \leq 1$) for the same event, respectively. One would want to reach to a consensus based on these two different opinions. This goal can be achieved by incorporating various functions that operate on input probabilities. The $\mathrm{AND}$ operator is one such function:

$$\mathrm{AND}(p,q) = pq \qquad (15)$$

Another option could be the $\mathrm{OR}$ operator, defined as:

$$\mathrm{OR}(p,q) = p + q - pq \qquad (16)$$

If the opinion of one of the decision makers is superior to the other one, the $\mathrm{OR}$ operator can be easily modified by employing a weight parameter $\alpha$ ($0 \leq \alpha \leq 1$):

$$\mathrm{OR}_{\text{weighted}}(p,q) = p + (q - pq)\alpha \qquad (17)$$

The more interesting function in the context of current research is the $\mathrm{OR}$ operator, for two reasons: 1) fits much better in the problem of link prediction as it is less prone to variations of only one of the input probabilities, 2) the weighted form provides a parameter to control the superiority of one of the input opinions. We will return to the issue of fusion of decisions in the following sub-section when characterizing the link prediction model.

## The Multiplex Link Prediction Model

On these grounds, a model is suggested to predict probability of connection between node pairs in a layer of the multiplex network such as $T$ which incorporates information both from the layer itself and from some other auxiliary layer $A$. The similarity between two distinct nodes $x$ and $y$ is defined as:

$$SB_{xy}^{T,A} = -I(L_{xy}^T = 1 | S_i^T, S_j^A); \quad (x,y) \in S_i^T \cap S_j^A \tag{18}$$

where $I(L_{xy}^T = 1 | S_i^T, S_j^A)$ is the uncertainty of existence of a link between $(x,y)$ in the target layer when their target and auxiliary bin numbers are known. According to equation (4), we can write:

$$-I(L_{xy}^T = 1 | S_i^T, S_j^A) = -I(L_{xy}^T = 1) + I(L_{xy}^T = 1; S_i^T, S_j^A) \tag{19}$$

The first term in equation (19) can be derived by incorporating equation (2):

$$-I(L_{xy}^T = 1) = \log p(L_{xy}^T = 1) \approx \log(\tilde{S}_{xy}^T) \tag{20}$$

where $\tilde{S}_{xy}^T$ is the min-max normalized similarity score of the pair $(x,y)$ in target layer $T$, i.e. the probability of connection in target layer (without any knowledge on bins partitioning) is estimated with similarity in that same layer, intuitively. The second term in equation (19) is the mutual information of $(x,y)$ being connected in the target layer and belonging to $S_i^T$ and $S_j^A$ bins; which is estimated as follows:

$$I(L_{xy}^T = 1; S_i^T, S_j^A) \approx I(L^T = 1; S_i^T, S_j^A) \tag{21}$$

Equation (21) propounds the view that a group of node pairs dwelling in known target and auxiliary bins can be looked at similarly. To be more specific, if the goal is to obtain the mutual information between the event that $(x,y)$ are connected and the event that it resides in both $S_i^T$ and $S_j^A$, a possible workaround is to estimate it with the reduction in uncertainty of connection of *any* node pair due to which bins (target and auxiliary) it belongs to. Thus, according to equation (4), we proceed by expanding the right-hand side of equation (21):

$$I(L^T = 1; S_i^T, S_j^A) = I(L^T = 1) - I(L^T = 1 | S_i^T, S_j^A) \tag{22}$$

The term $I(L^T = 1)$ in equation (22) is the self-information of that a randomly chosen node pair is linked in target layer $T$. Clearly, $I(L^T = 1)$ is the same for every node pair in the multiplex network; therefore, it does not affect the scoring (node pairs ranking), and it can be safely neglected. Thus, to carry out the model specification, $I(L^T = 1 | S_i^T, S_j^A)$ needs to be calculated; which is the conditional self-information of that a randomly chosen node pair is linked in layer $T$ when the pair's state of binning in target and auxiliary layer is known. Using equation (2) we have $I(L^T = 1 | S_i^T, S_j^A) = \log p(L^T = 1 | S_i^T, S_j^A)$. On the basis of our discussion on fusion of decisions, the probability $p(L^T = 1 | S_i^T, S_j^A)$ for any randomly selected node pair $(x,y)$ which is a member of $S_i^T \cap S_j^A$ is estimated by incorporating $p_{\text{intra}}(S_i^T)$ i.e. intra-layer connection probability in target layer $T$ and $p_{\text{trans}}^T(S_j^A)$ i.e. trans-layer connection probability in $T$ w.r.t auxiliary layer $A$. Therefore, similar to equation (17), the weighted $\text{OR}$ of intra and trans-layer connection probabilities concludes in:

$$p(L^T = 1 | S_i^T, S_j^A) = p_{\text{intra}}(S_i^T) + \left[ p_{\text{trans}}^T(S_j^A) - p_{\text{intra}}(S_i^T) p_{\text{trans}}^T(S_j^A) \right] \frac{j-1}{b_A} \tag{23}$$

$$= \left[ P_{\text{est}}^{T,A} \right]_{ij}$$

The term $\frac{j-1}{b_A}$ reflects the extent to which the decision of the auxiliary layer is considered in fusion of decisions. When similarity is very low in the auxiliary layer, the formula neglects its' decision. On the contrary, high similarity in the auxiliary layer applies $\text{OR}$ operator to decisions of the target and auxiliary layers. It should be further noticed that the trans-layer connection probability can be divided for connected and unconnected node pairs in the auxiliary layer according to equation (13) and (14), respectively.

To put it altogether, we incorporate equations (13) and (14) into (23). Then, plugging equation (23) into equation (18) results in the final scoring scheme. Thus, SimBins similarity score of a node pair $(x,y)$ in target layer $T$ with the aid of auxiliary layer $A$ where $(x,y) \in S_i^T \cap S_j^A$; $i \in \{1,...,b_T\}, j \in \{1,...,b_A\}$ and $T,A \in \{1,...,M\}; T \neq A$ is (empirical values of intra and trans-layer connection probabilities are used):

$$SB_{xy}^{T,A} = \begin{cases} \log(\tilde{s}_{xy}^T) + \log\left(p_{\text{intra}}(S_i^T) + \left[\tilde{p}_{\text{trans}}^T(S_j^A \cap E^A) - p_{\text{intra}}(S_i^T)\tilde{p}_{\text{trans}}^T(S_j^A \cap E^A)\right]\frac{j-1}{b_A}\right) & ; (x,y) \in E^A \\ \log(\tilde{s}_{xy}^T) + \log\left(p_{\text{intra}}(S_i^T) + \left[\tilde{p}_{\text{trans}}^T\left(S_j^A \cap (U - E^A)\right) - p_{\text{intra}}(S_i^T)\tilde{p}_{\text{trans}}^T\left(S_j^A \cap (U - E^A)\right)\right]\frac{j-1}{b_A}\right) & ; (x,y) \in U - E^A \end{cases}$$
(24)

To summarize the whole method, the corresponding pseudo-code is given in Algorithm 1. Now that our multiplex scoring model is complete, we will proceed by evaluating the method on the datasets introduced earlier.

**Algorithm 1 Scoring node-pairs by SimBins.**

**input** : Multiplex network $G(V, E^T, E^A; E^T, E^A \in U = V \times V)$
**output:** Similarity scores for test set based on the proposed method

$T$ is the target layer and $A$ is the auxiliary layer;
Number of bins is initialized to $b$;
Number of training iterations is initialized to $iters$;

**for** $iteration \leftarrow 1$ **to** $iters$ **do**
  $E^T_{train}, E^T_{test} \leftarrow$ DivideNetToTrainTest($E^T$, $TrainingRatio$);
  $Z^T_{train} \leftarrow$ Sample($U - E^T_{train}, size = 2|E^T_{train}|$);
  $Z^T_{test} \leftarrow$ Sample($U - E^T_{test}, size = 2|E^T_{test}|$);
  $U' \leftarrow E^T_{train} \cup Z^T_{train}$;
  **foreach** node-pair $(x,y)$ in $U'$ **do**
    $sim^T_{xy} \leftarrow$ BaseSimilarity($x,y,T$);
    $sim^A_{xy} \leftarrow$ BaseSimilarity($x,y,A$);
  **end**
  $S^T_1, S^T_2, ..., S^T_b \leftarrow$ EqualDepthBinning($U', S^T$);
  $S^A_1, S^A_2, ..., S^A_b \leftarrow$ EqualDepthBinning($U', S^A$);
  **foreach** partition $S^T_i$ **do**
    Compute $p_{intra}(S^T_i)$ according to equation (10);
  **end**
  **foreach** partition $S^A_j$ **do**
    Compute $p^T_{trans}(S^A_j \cap E^A)$ and $p^T_{trans}(S^A_j \cap (U - E^A))$ according to equation (13) and (14), respectively;
  **end**
  **foreach** 2d-bin $(S^T_i, S^A_j)$ **do**
    Compute $[P^{T,A}_{est}]_{ij}$ according to equation (23);
  **end**
  **foreach** node-pair $(x,y)$ in $E^T_{test} \cup Z^T_{test}$ residing in 2d-bin $(S^T_i, S^A_j)$ **do**
    $SB^{T,A}_{x,y} \leftarrow$ according to equation (24), using equation (13) or (14) based on linkage of $(x,y)$ in $A$;
  **end**
**end**
**return** $[SB^{T,A}]$;

## Experimental Results

The link prediction performance on 8 different datasets and a total of 24 network layers has been reported in Table 2. The evaluation metric is the average AUC over 100 training phases (iterations) with train ratio set to 90% as described in 'Evaluation Method' section. Three base measures has been incorporated i.e. RA, CN and ACT that were explained in 'Base Similarity Measures' section. SimBins ( $SB_T^A = SB^{T,A}$) is compared with scoring based on similarity in the target layer ($S_T$) and the simple addition of similarity scores of the target and auxiliary layer ($S_T + S_A$).

**Table 2 Average AUC over 100 iterations for the networks under study.** Each row shows the performance of link prediction methods on a duplex subset of a multiplex network. Duplexes are grouped by their corresponding network (dataset). Columns show the average AUC over 100 iterations for prediction methods $S_T$ (similarity score of only the target layer), $S_T + S_A$ (addition of similarity scores of the target and auxiliary layer), $SB_T^A$ (SimBins score) grouped by base similarity measure used.

| | | | RA | | | CN | | | ACT | | |
|---|---|---|---|---|---|---|---|---|---|---|---|
| | Target Layer | Auxiliary Layer | $S_T$ | $S_T + S_A$ | $SB_T^A$ | $S_T$ | $S_T + S_A$ | $SB_T^A$ | $S_T$ | $S_T + S_A$ | $SB_T^A$ |
| AT | Air | Train | 82.6 | 90.2 | **91.3** | 81.5 | **85.8** | 85.5 | **88.3** | 86.4 | 88.1 |
| | Train | Air | 82.9 | **83.8** | **83.8** | 82.9 | 83.1 | **83.8** | 80.0 | 80.5 | 81.3 |
| C. ELEGANS | Electric | Chem-Mono | 70.7 | **79.3** | **79.3** | 70.1 | 78.0 | **78.8** | 64.9 | 66.4 | 69.4 |
| | | Chem-Poly | 71.2 | 84.8 | **86.0** | 70.6 | 83.3 | **85.6** | 65.4 | 68.1 | 71.9 |
| | Chem-Mono | Electric | 76.4 | 77.2 | **77.5** | 75.3 | 75.9 | **76.6** | 68.6 | 68.9 | 71.5 |
| | | Chem-Poly | 76.1 | 87.2 | **87.7** | 75.9 | 85.4 | **88.2** | 68.9 | 74.0 | 83.4 |
| | Chem-Poly | Electric | 85.8 | 85.8 | **86.2** | 83.9 | 83.9 | **84.2** | 72.4 | 71.9 | 73.2 |
| | | Chem-Mono | 85.8 | 86.8 | **87.3** | 83.8 | 85.4 | **85.9** | 72.3 | 72.8 | 79.6 |
| DM | Suppressive | Additive | **76.8** | 76.1 | **76.8** | **76.7** | 75.9 | 76.4 | **81.0** | 74.1 | 75.7 |
| | Additive | Suppressive | 74.0 | 73.8 | **74.2** | **74.1** | 73.3 | **74.1** | **74.4** | 70.8 | 68.8 |
| HB | Structure | Function | 91.7 | 91.4 | **93.5** | 90.7 | 90.0 | **92.9** | 74.4 | 68.1 | 78.1 |
| | Function | Structure | 86.2 | 89.4 | **90.3** | 85.4 | 88.7 | **90.1** | 68.8 | 72.1 | 78.3 |
| PHYSICIANS | Advice | Discuss | 72.1 | **82.5** | **82.5** | 71.5 | **82.2** | **82.6** | 50.0 | 65.9 | 74.9 |
| | | Friendship | 72.4 | **78.6** | **78.6** | 71.9 | 78.3 | **78.5** | 51.0 | 59.2 | 63.3 |
| | Discuss | Advice | 74.5 | **80.8** | **80.8** | 75.2 | 81.3 | **81.7** | 52.1 | 62.0 | 70.4 |
| | | Friendship | 74.2 | **81.0** | **81.0** | 74.8 | 80.5 | **80.7** | 52.3 | 62.4 | 67.9 |
| | Friendship | Advice | 70.2 | 77.9 | **78.2** | 69.8 | 76.9 | **77.3** | 56.2 | 58.0 | 63.5 |
| | | Discuss | 69.9 | 81.8 | **82.1** | 70.1 | 82.0 | **82.4** | 56.4 | 65.2 | 71.3 |
| NTN | Communi. | Financial | **85.0** | 84.8 | 84.2 | **81.8** | 81.7 | **81.8** | 75.6 | 63.1 | 72.3 |
| | | Operation | 84.9 | 85.9 | **87.8** | 84.0 | 83.9 | **88.4** | 74.5 | 67.1 | 78.9 |
| | | Trust | 85.7 | 85.0 | **87.5** | 83.5 | 80.8 | **84.6** | 75.0 | 72.2 | 80.2 |
| | Financial | Communi. | 89.0 | **91.7** | 89.6 | **91.0** | 82.5 | 90.5 | 49.0 | 35.3 | 66.3 |
| | | Operation | **91.8** | 86.9 | 91.0 | 91.5 | 66.2 | **92.7** | 49.7 | 51.1 | 73.6 |
| | | Trust | 89.2 | 94.2 | **96.1** | 89.5 | 84.7 | **95.9** | 49.8 | 41.1 | 74.5 |
| | Operation | Communi. | 98.2 | 98.2 | **98.7** | 97.1 | 97.4 | **97.8** | 66.9 | 68.7 | 80.2 |
| | | Financial | **98.3** | 97.9 | **98.3** | 97.2 | 97.2 | 97.2 | 66.6 | 59.0 | 74.3 |
| | | Trust | 98.0 | 95.9 | **98.3** | 97.1 | 94.6 | **97.1** | 67.6 | 64.8 | 77.8 |
| | Trust | Communi. | 88.8 | 92.3 | **92.7** | 87.8 | 91.4 | **92.2** | 78.3 | 80.1 | 87.7 |
| | | Financial | **88.3** | 88.0 | **88.3** | **88.3** | **88.3** | **88.3** | 78.8 | 67.3 | 80.6 |
| | | Operation | 88.8 | 88.0 | **91.5** | 87.4 | 86.0 | **90.9** | 78.3 | 70.8 | 83.7 |
| LONDON TRANS | Tube | Overground | **53.4** | **53.4** | **53.4** | **53.3** | **53.3** | **53.3** | 52.9 | 46.0 | 50.9 |
| | | DLR | **53.7** | **53.7** | **53.7** | **53.3** | **53.3** | **53.3** | 53.4 | 49.5 | 46.6 |
| | Overground | Tube | **50.0** | **50.0** | **50.0** | **50.0** | **50.0** | **50.0** | 49.1 | 51.5 | 63.5 |
| | | DLR | 49.9 | 49.9 | **50.2** | 49.9 | 49.8 | 50.0 | 48.9 | **50.2** | 47.6 |
| | DLR | Tube | 53.2 | **54.2** | 49.6 | 52.0 | **52.9** | 50.0 | 60.1 | 61.6 | **65.7** |
| | | Overground | **52.9** | **52.9** | 50.0 | **52.8** | **52.8** | 50.5 | **59.9** | 58.1 | 47.8 |
| CS-AARHUS | Lunch | Facebook | 95.1 | 93.4 | **95.5** | 94.2 | 90.6 | **94.7** | **83.3** | 59.9 | 81.6 |
| | | Co-author | **94.7** | 94.5 | **94.7** | 93.6 | 93.7 | **93.7** | 83.1 | 56.1 | 81.5 |
| | | Leisure | **94.5** | 94.0 | **94.5** | 93.8 | 93.9 | **94.1** | 82.7 | 69.1 | 81.8 |
| | | Work | 94.7 | 94.8 | **95.9** | 94.4 | 93.5 | **95.4** | 84.0 | 81.5 | **87.6** |
| | Facebook | Lunch | **93.7** | 90.4 | 93.4 | **92.6** | 90.4 | **92.6** | 42.1 | 49.9 | 78.9 |
| | | Co-author | **92.5** | 92.3 | **92.5** | 92.0 | 91.8 | **92.0** | 41.8 | 47.6 | 72.0 |
| | Co-author | Lunch | 69.9 | **92.2** | 91.1 | 71.0 | 92.1 | **92.9** | 43.0 | 60.0 | 76.8 |
| | | Facebook | 70.5 | 70.7 | **73.5** | 72.6 | 70.2 | **75.3** | 39.2 | **61.7** | 60.6 |
| | Leisure | Lunch | 81.1 | **90.2** | **90.2** | 82.2 | 89.9 | **91.0** | 59.3 | 75.0 | 80.8 |
| | Work | Lunch | 88.7 | 90.8 | **91.5** | 86.7 | 89.8 | **90.3** | 70.3 | **82.9** | 81.5 |
| | AVERAGE | | 81.5 | 84.3 | **84.9** | 81.0 | 82.5 | **84.5** | 65.8 | 65.5 | 74.8 |

For each base measure, the highest average AUC is shown in bold and, for each duplex (row), the highest AUC among all of the methods (independent from the base measure) is highlighted with an underscore. SimBins dominates other two methods and proves to be an effective multiplex link prediction method due to several reasons: ***(i)*** Most of the time, SimBins is superior to the other methods (i.e. bold entries). ***(ii)*** In a large fraction of duplexes (27 of 46), the overall best mean AUC belongs exclusively to SimBins (in 12 other duplexes, SimBins achieves the best performance alongside another method, non-exclusively) ***(iii)*** SimBins performs better than the single-layer method (or $S_T$) in most of the cases whereas for similarities addition method ($S_T + S_A$) this is less frequently observed; meaning our method is capable of using other layer's information effectively. And, $SB^{T,A}$ is more robust against deceptive signals compared to $S_T + S_A$. Consider Drosophila in Table 2 for example. The slightly negative correlation between similarity in the auxiliary layer (Suppressive) and connection probability in the target layer (Additive), as previously discussed on (Fig 2-d), has caused performance reduction for $S_T + S_A$ whereas SimBins still performs as good as —if not better than— $S_T$. A similar outcome can be observed for NTN and London Transport, more clearly when ACT is used as the base similarity measure. In CS-Aarhus, where Facebook is the target layer, both $S_T$ and $S_T + S_A$ perform even worse than random scoring (expected $50\%$ AUC) while SimBins keeps the performance up about $70-80\%$. As the last row in Table 2 dictates, the average mean AUC of SimBins is higher than both other methods, no matter the choice of base measure.

There exist occasions in which SimBins cannot improve the link prediction performance compared to the base similarity measure. Specifically, Drosophila which the absence of inter-layer correlation as discussed earlier is the underlying reason. And, in London Transport, node multiplexity is far too low as shown in Table 1; consequently, very few nodes are shared among different layers that makes utilization of structural similarities between layers a hard task. Evidently, the AA scores of Overground and DLR layers in London Transport are almost all zeros, hence is the $50\%$ AUC.

The above discussion holds true for Adamic-Adar [9] and Preferential Attachment [8] similarity measures, as we have performed similar experiments which led to resembling results, but we have avoided bringing the corresponding details for the sake of brevity.

Interestingly, the results appear to suggest that choosing RA as the base similarity measure, leads to the best overall performance in most of the multiplex networks.

## Complexity Analysis

Consider a duplex network $G(V, E^{[1]}, E^{[2]}; E^{[i]} \subseteq V \times V), m_i = |E^{[i]}| \ \forall i \in \{1, 2\}$ where layer $1$ is the target, and layer $2$ is the auxiliary layer. Let $O(\theta)$ be a representative of computational complexity for the base similarity measures. The similarity of node pairs in both layers is needed for subset $U'$ of $U = V \times V$ as formulated in 'Partitioning Node Pairs (Binning)' section. Therefore, the computing complexity of measuring similarities is $O(\sum_{i=1,2} \theta m_i)$. Partitioning $U'$ into equal-depth bins requires sorting of similarities, consequently it would have complexity of $O(\sum_{i=1,2} m_i \log m_i)$. Total estimation complexity of intra-layer and trans-layer connection probabilities is $O\left(\sum_{i=1,2} m_i b_i\right)$ where $b_i$ is the number of bins in corresponding layer. And, estimation of probability of connection in all 2d-bins according to equation (23) would be of order $O(b_1 b_2)$ which is negligible w.r.t

bounded number of bins. Accordingly, the total computational complexity of scoring a node pair in SimBins would be $O(m \log m)$ where $m$ is in the same order as $m_1, m_2$. if the sparsity of multiplex layers are comparable. This tolerable computing complexity indicates that SimBins can be scaled for usage in large networks.

Notice that for obtaining a full ranking of propensity of links, SimBins, like the majority of link prediction algorithms would need at least $O(n^2); n = |V|$ computations which is not easily scalable to very large networks without pruning the $n^2$ space. To be specific, for a full ranking, SimBins would have a computing complexity of $O(\theta n^2 + m \log m)$ in which $O(\theta n^2)$ is the dominating term in real-networks; meaning that SimBins imposes minor overhead to the base similarity measures.

## Discussion and Conclusions

In this manuscript, we explored the intra-layer and trans-layer connection probabilities in multiplex networks and verified that in many real multiplex networks, connection probability in some layer is correlated with similarity in another layer of the same multiplex. Subsequently, we developed a link prediction model by incorporating information theory concepts for characterizing intuitions gather from observed data.

The proposed method works on a pair of multiplex's layers i.e. a duplex. Different ideas can be conducted to extend it to use multiple layers' topology for link prediction. Considering a target layer $T$ and auxiliary layers $A_1, ..., A_M$, the simplest idea is to add up the SimBins scores for each possible layer pairs, symbolically $\mathrm{SB}^{T, \{A_1, ..., A_M\}} = \sum_{i=1}^{M} \mathrm{SB}^{T, A_i}$ where $\mathrm{SB}^{T, A_i}$ is computed according to equation (24). The other –not as straightforward as previous– idea is to compose and study bins of more than two dimensions. This extension, although more systematic, might suffer from heavy sparsity of samples (imagine node pairs residing in 3d-bins).

Eventually, SimBins is compared with a single-layer method and a multiplex method on 8 multiplexes; (1) base similarity measure in the target layer and (2) simple addition of similarities in target and auxiliary layers, respectively. It is shown that SimBins outperforms the other two methods in most cases (up to 30% mean AUC boost in some cases). Besides, it performs worse than target similarity very rarely and is more robust to deceptive signals compared to simple addition of similarities. It is mentioned that in some networks, such as London Transport and Drosophila, SimBins seems to be unprofitable as a result of massively condensed node pairs similarity distribution and negative inter-layer correlations.

It is shown that SimBins imposes negligible computation overhead to the base similarity measures. The idea of using an equal-width strategy for partitioning node pairs leads to even more efficiency due to its $O(m)$ complexity (instead of $O(m \log m)$ in equal-depth binning), although the accuracy of prediction might be affected.

## Acknowledgements

We express our thanks to Dr. Behnam Bahrak for reviewing the manuscript and providing helpful comments and insights.